\documentclass[showpacs,aps,amsmath,letterpaper,nofootinbib,reprint,showkeys,superscriptaddress,longbibliography,prl]{revtex4-2}
\usepackage{dcolumn}    
\usepackage{bm}         
\usepackage{ifpdf}
\usepackage{amssymb,lineno,amsfonts}
\usepackage{graphicx}   
\usepackage{bbm}
\usepackage{mathrsfs}
\usepackage{mathtools}
\usepackage{upgreek}

\usepackage{epstopdf}
\usepackage{setspace}
\usepackage{hyperref}
\usepackage{cleveref}
\usepackage{bbold}
\usepackage[matrix,frame,arrow]{xypic}
\usepackage{float}
\usepackage{soul}
\usepackage{tikz}
\usepackage{natbib}
\usepackage{color} 
\usepackage{subfigure}
\usepackage{bbold}
\usepackage{tikz}
\usepackage{amsthm}
\usepackage{academicons}
\usepackage[english]{babel}

\definecolor{med-blue}{RGB}{25,25,112} 
\hypersetup{colorlinks, linkcolor={blue},citecolor={blue}, urlcolor={blue}}

\newcommand{\Eqref}[1]{(\ref{#1})}

\newcommand{\expo}[1]{\mathrm{e}^{#1}}

\newcommand{\outpr}[2]{\vert{#1}\rangle\langle{#2}\vert}
\newcommand{\proj}[1]{\vert{#1}\rangle \langle{#1}\vert}
\newcommand{\Ket}[1]{\vert{#1}\rangle}

\newcommand{\dnorm}[1]{\vert\vert{#1}\vert\vert}

\usepackage{orcidlink}

\begin{document}
\title{Direct Experimental Observation of Quantum Mpemba Effect without Bath Engineering}
\author{Arijit Chatterjee {$\ddagger$} \orcidlink{0009-0000-8191-1882}}
\email{arijitchattopadhyay01@gmail.com}
\affiliation{Department of Physics and NMR Research Center,\\
Indian Institute of Science Education and Research, Pune 411008, India}
\author{Sakil Khan}
\email{sakil.khan@students.iiserpune.ac.in}
\affiliation{Department of Physics,\\
Indian Institute of Science Education and Research, Pune 411008, India}
\author{Sachin Jain}
\email{sachin.jain@iiserpune.ac.in}
\affiliation{Department of Physics,\\
Indian Institute of Science Education and Research, Pune 411008, India}
\author{T. S. Mahesh}
\email{tsmahesh@gmail.com}
\affiliation{Department of Physics and NMR Research Center,\\
Indian Institute of Science Education and Research, Pune 411008, India}
\def\thefootnote{$\ddagger$}\footnotetext{Contact author : \color{blue}{arijitchattopadhyay01@gmail.com}}\def\thefootnote{\arabic{footnote}}

\begin{abstract}
{The quantum Mpemba effect refers to the phenomenon of a quantum system in an initial state, far away from equilibrium, relaxing much faster than a state comparatively nearer to equilibrium. We experimentally demonstrate that this highly counterintuitive effect can occur naturally during the thermalization of quantum systems. Considering dipolar relaxation as the dominant decoherence process, we theoretically derive the conditions that can lead to the Mpemba effect in nuclear spins. After experimentally preparing nuclear spin states dictated by those conditions, we observe the occurrence of the Mpemba effect when they are left to thermalize without any external control. We also experimentally observe the genuine quantum Mpemba effect during thermalization of nuclear spins. Our results establish that both these effects are natural in thermalization of quantum systems, and may show up without the need for any bath engineering.}
\end{abstract}

\maketitle

\begin{figure}
\includegraphics[width=8.6cm,  clip=true, trim={0cm 11.7cm 8.8cm 2cm}]{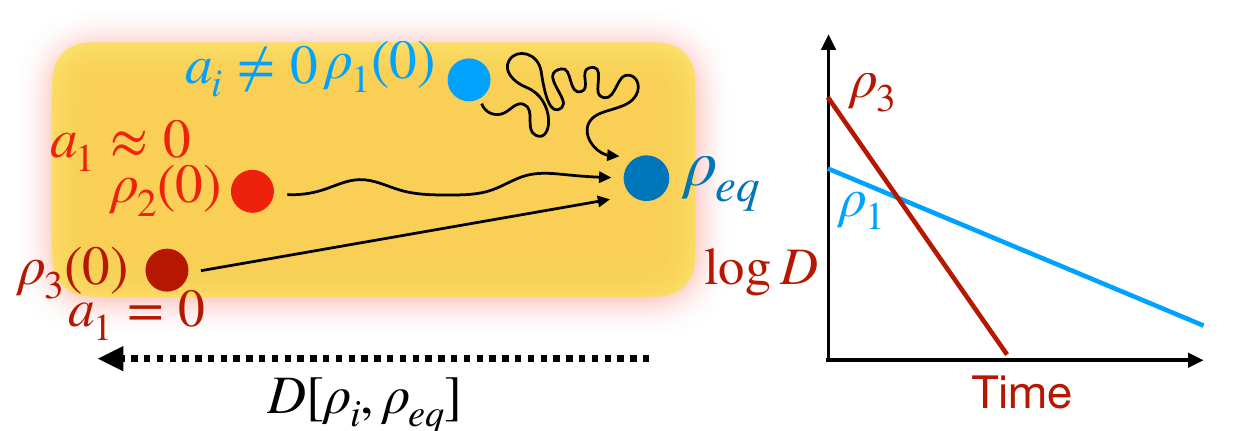}
\caption{(left) The Mpemba effect occurs when a generic state $\rho_1$, overlapping ($a_i$ is the overlap with $i$'th decay mode with $i=1$ being the slowest) with all decay modes, relaxes much slowly than a state that is far away from the steady state $\rho_{eq}$ w.r.t a metric $D$, but have near zero ($\rho_2$) or zero ($\rho_3$) overlap with the slowest decay mode. (right) A typical plot of the observation of Mpemba effect with respect to metric $D$, where the crossing indicates that the far state has overtaken the nearer state and reached the equilibrium faster.} 
\label{fig:grpabs}
\end{figure}

\emph{Introduction}--- Mpemba effect refers to a situation where an initially hotter system, when brought into contact with a cold bath, reaches the thermal equilibrium more rapidly than an initially cooler system. Even though this rather surprising phenomenon finds its mentions in various texts throughout history  \cite{alma991000934559708966,999697569602121,Bacon1962,bacon1902novum,groves2009now}, the first systematic studies by Mpemba and Osborne \cite{E_B_Mpemba_1969}, as well as Kell \cite{10.1119/1.1975687} in 1969 sparked huge interest in the community, followed by a series of  investigations on a variety of systems \cite{Greaney2011,Ahn2016,PhysRevLett.119.148001,Keller_2018,Hu2018,doi:10.1073/pnas.1819803116,chaddah2010overtakingapproachingequilibrium}. Despite being studied \cite{PhysRevLett.129.138002,Holtzman2022,PhysRevLett.131.017101,PhysRevLett.132.117102,PhysRevLett.124.060602,walker2023mpembaeffecttermsmean,walker2023optimaltransportanomalousthermal,bera2023effectdynamicsanomalousthermal,PhysRevE.109.044149,PhysRevLett.134.107101,teza2025speedupsnonequilibriumthermalrelaxation} over many decades, the very conditions for the occurrence of this effect continues to be questioned \cite{Vynnycky2010,Burridge2016,burridge2020observing,Bechhoefer2021} and lacks uniform agreement. First details-independent uniform understanding of Mpemba effect came  by trying to find its presence when an out-of-equilibrium system approaches equilibrium via Markovian evolution \cite{doi:10.1073/pnas.1701264114}. Considering a distance function $D$, which measures how far the system is from the equilibrium, Mpemba effect can be formally defined as a scenario when the system that is far from equilibrium relaxes quickly than a system starting relatively nearer to equilibrium. At long times, the dynamics of such a system is determined by its slowest and second slowest decay modes. Mpemba effect can show up if the far state has much smaller overlap with the slowest decay mode in comparison to the nearer state. The far state can then effectively relax quickly through the second slowest decay mode, while the near state undergoes a slow relaxation through the slowest decay mode (see Fig.~\ref{fig:grpabs}). A stronger version of it, known as strong Mpemba effect, \cite{PhysRevX.9.021060}  happens when the overlap between the far state and the slowest decay mode becomes exactly zero. This transparent explanation of Mpemba effect in Markovian setting resolved many debates regarding its origin, and also got experimental confirmation \cite{Kumar2020,doi:10.1073/pnas.2118484119}, even though certain observations of Mpemba effect still lies outside the Markovian setting and call for system specific explanations \cite{VYNNYCKY2015243,10.1119/1.18059,ESPOSITO2008757,10.1119/1.2996187,C4CP03669G,Jin2015,https://doi.org/10.1002/crat.2170230702}.    

The understanding of Mpemba effect in Markovian systems prompted investigations regarding its existence in the quantum realm. When a quantum system relaxes to equilibrium by interacting with a sufficiently large and weakly coupled reservoir, its evolution is governed by a GSKL master equation \cite{breuer2002theory,lidar2020lecturenotestheoryopen}, whose solution can be written down as a sum of the decay modes much like the classical case \cite{doi:10.1073/pnas.1701264114}. A generic initial quantum state will have overlap with all the decay modes and hence undergoes a slow relaxation  with the timescale dictated by the slowest decay mode. An exponentially accelerated relaxation \cite{PhysRevLett.127.060401,bao2025acceleratingquantumrelaxationtemporary,PhysRevA.106.012207,PhysRevE.108.014130} can be obtained by applying a suitable unitary operator to the state such that its overlap with the slowest decay mode gets minimized, or become zero (see Fig.~\ref{fig:grpabs}).  In this process, if the action of the unitary also pushes the system much farther from the unique steady state but still it relaxes faster than a nearer state, then it is referred to as the quantum Mpemba effect
\cite{PhysRevLett.127.060401,bao2025acceleratingquantumrelaxationtemporary,PhysRevA.106.012207,PhysRevE.108.014130,PhysRevResearch.3.043108,PhysRevResearch.6.033330,PhysRevLett.133.136302,PhysRevLett.131.080402,PhysRevA.110.022213,PhysRevLett.133.140404,Ares2025,PhysRevA.110.042218,Longhi2025mpembaeffectsuper,PhysRevA.111.022215,zhao2025noiseinducedquantummpembaeffect,chang2024imaginarytimempembaeffectquantum,10.1063/5.0234457,kheirandish2024mpembaeffectquantumoscillating,PhysRevB.111.125404,PhysRevLett.134.220402,FURTADO2025170135}. 
Even though there have been reports of Mpemba like effects beyond Markovian
settings  \cite{PhysRevLett.133.010402,5d6p-8d1b,PhysRevLett.134.220403,PhysRevLett.133.140405,PhysRevLett.133.010401,Caceffo_2024,PhysRevB.111.104312,Chalas_2024}, in this article we will primarily focus on the  Markovian evolution.         

Experimental demonstration of quantum Mpemba effect in markovian setting have recently been achieved using trapped ions \cite{PhysRevLett.133.010403,Zhang2025}, where the relaxation is realized by coupling the system states with one or more metastable states. Since the system relaxes through engineered decay channels in the respective experiments, it does not thermalize in the end. Therefore, corresponding to the classical case, the key question of whether quantum Mpemba effect can occur or not in natural thermalization of quantum systems remains open. In this work we provide an affirmative answer to this question by observing quantum Mpemba effect in thermalization of nuclear spins. Considering dipolar relaxation as the dominant decoherence process in nuclear spins, we derive the conditions which can lead to the occurrence of quantum Mpemba effect. In the experiments, we prepare nuclear spin states satisfying those conditions and then allow them to relax without applying any pulses to avoid any external influence to the natural thermalization. The fact that they end up in the thermal state is also experimentally confirmed through tomography. We also experimentally demonstrate that even `genuine quantum Mpemba effect' \cite{PhysRevLett.133.140404}, where a system with higher free energy attains equilibrium faster than the one with lower free energy, can happen naturally during thermalization. Our results establishes that both quantum Mpemba effect, and `genuine quantum Mpemba effect' are  natural phenomena that can occur as quantum system thermalises without the requirement of any bath engineering or specially induced relaxation.       

\emph{Thermalization of Nuclear spins}---We consider an ensemble of homonuclear two spin $1/2$ systems in the presence of a strong magnetic field $B_0\,\hat{z}$. The spin Hamiltonian of such systems reads \cite{cavanagh1996protein,levitt2008spin,keeler2011understanding,Wong2014}
\begin{equation}
\frac{1}{\hbar}\mathcal{H} = \left(\omega_0 - \frac{\Delta}{2}\right) I_{1z} +\left(\omega_0 + \frac{\Delta}{2} \right) I_{2z}  + 2\pi J I_{1z} \, I_{2z}, \label{eq:nmr_Ham}
\end{equation}
where  $I_{kl}$ is the spin angular momentum operator of the $k$'th spin along $\hat{l}$ direction. 
In the presence of a  strong field, the Larmor frequency $\omega_0/(2\pi)$ remains the dominant parameter (few hundreds of MHz) while resonance offset ($\Delta/(2\pi)$) and scalar coupling $(J)$ can be treated as perturbations ($\approx 0.1 -  1$ KHz). The sample molecules, with nuclear spin Hamiltonian of Eq.~\Eqref{eq:nmr_Ham}, are dissolved in some  solvent to prepare a dilute isotropic solution, which is kept at temperature $T$. Initially, the nuclear spins rest in thermal equilibrium state 
\begin{gather}
\rho^{th} = \expo{-\mathcal{H}/k_B T}/Z(T) \approx \left[ \mathbbm{1} + 2\epsilon \left( I_{1z} + I_{2z} \right)\right]/4, \label{eq:thstate}
\end{gather}
where $Z(T)=\text{tr}[\exp(-\mathcal{H}/k_B T)]$ is the partition function with Boltzmann constant $k_B$. Since the temperature $T$ is usually very high ($k_BT \gg \omega_0 \hbar$), we only consider up to the first order in $\epsilon = -\omega_0\hbar/(2k_BT)$. The thermal state populations thus differ from maximally mixed state $\mathbbm{1}/4$ by first order in $\epsilon$ : using trace distance \cite{barnett2009quantum,10268381,Nielsen_Chuang_2010} $D(\rho_A,\rho_B) = |\rho_A-\rho_B|/2$, with $|X| = \text{tr}\sqrt{X^{\dagger}X}$, to measure the distance between two states $\rho_A$ and $\rho_B$, we notice $D(\rho^{\text{th}}, \mathbbm{1}/4) \propto \epsilon \approx 10^{-5}$. However, modern day NMR spectrometers are sensitive enough to amplify the signal coming from this small polarization $(\epsilon)$ with sufficient noise suppression
\cite{doi:10.1073,PhysRevA.62.052314}.

When the nuclear spin state is perturbed from equilibrium, it always comes back to the thermal state $\rho^{th}$ of Eq.~\Eqref{eq:thstate} in some characteristic timescale ($T_1$). The irreversible dynamics of this thermalization process can be described by a GSKL master equation \cite{BENGS2020106645,breuer2002theory}. In the interaction frame of $U_0(t)=\exp(i\omega_0(I_{1z}+I_{2z})t)$, the density operator and the Hamiltonian transform as $\rho \rightarrow \widetilde{\rho}(t)=U_0(t) \rho(t)U_{0}^{\dagger}(t)$, and $\mathcal{H}\rightarrow \widetilde{\mathcal{H}} = U_0(t) \mathcal{H}\,U_{0}^{\dagger}(t) - i\,U_0(t)\dot{U_{0}^{\dagger}}(t)$ (the upper dot $\dot{(\,)}$ means partial derivative w.r.t. $t$), respectively. The master equation in this frame takes the following form (see Appendix B for derivation)
\begin{gather}
\dot{\widetilde{\rho}}(t) = -i\left[\widetilde{\mathcal{H}}, \widetilde{\rho}(t) \right] +  \sum_{m=-2}^{2} \mathcal{K}(m\omega_0)\,\Gamma\left[T_{2m},T^{\dagger}_{2m}  \right] \widetilde{\rho}\,(t), \nonumber \\
\text{where,}~~\Gamma[A,B](\cdot)=A(\cdot)B-\frac{1}{2}\{BA,(\cdot)\},
\label{eq:GSKL}
\end{gather}  
and $\{T_{2m}\}$ for $m\in\{-2,-1,0,1,2\}$ are the rank-$2$ irreducible spherical tensor operators \cite{Kimmich1997,sakurai1986modern} (see Appendix A,B for details). We have considered the dipolar relaxation \cite{abragam1961principles,10.1063/1.469808,ernst1990principles,GOLDMAN2001160} here which is usually the most dominating relaxation mechanism in two spin systems. Assuming an exponential decay of two-time bath correlations beyond a characteristic time $\tau_c$, the spectral density function can be written as      
\begin{equation}
\mathcal{K}(x) = K(x)\,\expo{x\epsilon/\omega_0}, ~\text{where}~K(x)=\frac{12\,b^2 \tau_c}{5(1+x^2\tau_{c}^2)}
\end{equation}
is the spectral density computed from semiclassical theory \cite{mr-3-27-2022,abragam1961principles} (with $b$ being the dipolar coupling constant), which is multiplied by the temperature correction \cite{PhysRevLett.4.239,doi:10.1021/jp9919314} to ensure quantum detailed balance \cite{BENGS2020106645,PhysRevE.98.052104,PhysRevE.98.052104}. The formal solution of Eq.~\Eqref{eq:GSKL} reads $\widetilde{\rho}(0)=\exp(\mathcal{L}t)\rho(0)$, where the generator $\mathcal{L}$ is called the Liouvillian \cite{Nakano2011,PhysRevLett.130.230404} and the spectrum of $\mathcal{L}$ ultimately determines the possible occurrence of Mpemba effect. 

For small  molecules dissolved in liquid solvents at ambient temperatures ($T \approx 290-340$ K), we observe extremely fast random diffusion motion causing the bath correlation time to be very small \cite{cavanagh1996protein,levitt2008spin} in comparison to the dynamics under $\mathcal{H}_0$, which is called `extreme narrowing' $\omega_0\,\tau_c \ll 1$. Under this condition, we get $K(m\omega_0)\approx K_0 = 12b^2 \tau_c/5$. This, and the fact that experiments are done at high temperatures ($\mathcal{O}(\epsilon^2)\approx 0$), allow us to write the spectral density function as $\mathcal{K}(m\omega_0)\approx K_0 (1+m\,\epsilon)$.  

\emph{Eliminating the effect of dephasing}--- The Zeeman field $B_0$ is usually slightly inhomogeneous along the length of the solution ($\approx 4-6$ cm). As the sample molecules undergo random diffusion motion across that length, their spin states experience a randomly fluctuating magnetic field along $\hat{z}$ and gets quickly decohered. This process, known as inhomogeneous dephasing, generally adds to the dipolar relaxation mentioned in Eq.~\Eqref{eq:GSKL}. As we are interested in investigating quantum Mpemba effect in the natural thermalization of nuclear spins caused by their interaction with the thermal bath, we would like to eliminate the effect of this  inhomogeneous dephasing process.

To do so, we first note that the two-spin density operator can be expanded in a basis $\{\xi_i\}_{i=1}^{16}$, which breaks up into three blocks $\{\xi_i\} = \{\xi_i^{(0)}\}_{i=1}^{6}\,\cup\,\{\xi_i^{(1)}\}_{i=1}^{8}\,\cup \{\xi_i^{(2)}\}_{i=1}^{2}$ such that the elements of $\{\xi_i^{(m)}\}$ form the $m$-quantum block satisfying the commutation relation $[\xi_{i}^{(m)},I_{1z}+I_{2z}]=m \, \xi_{i}^{(m)}$. The elements of density operators in the 1- and 2-quantum blocks suffer maximum inhomogeneous dephasing as they pick up larger random phases under a fluctuating magnetic filed  along $\hat{z}$ in a given interval of time. However, the zero-quantum block (ZQB) mostly contains population terms ($\{\proj{00},\proj{01},\proj{10},\proj{11}\}$) which do not evolve under $\hat{z}$ field since $\Ket{0(1)}$ is the eigenstate of $\sigma_z$ with eigenvalue $1(-1)$. There are only two coherence terms ($\outpr{01}{10}, \outpr{10}{01}$) present in the ZQB which gets affected by the inhomogeneous dephasing. As the Liouvillian of Eq.~\Eqref{eq:GSKL} also admits this block diagonal structure $\mathcal{L}= 
\mathcal{L}_{0}\,\oplus \mathcal{L}_{1} \oplus \mathcal{L}_2$, where $\mathcal{L}_{i}$ is $\mathcal{L}$ projected in the $i$'th quantum block for $i=0,1$ or $2$, desnity matrix elements from different blocks do not get mixed up in the thermalization process. This remarkable symmetry, known as the Redfield kite structure \cite{abragam1961principles,cavanagh1996protein}, allows us to restrict the dynamics only in the ZQB so that the subsequent relaxation happens solely due to the natural thermalization of the nuclear spins, provided we ensure the coherence terms of ZQB to remain zero throughout.

\begin{figure}
\includegraphics[width=8.6cm, clip=true, trim={0cm 2.1cm 9.5cm 2.2cm}]{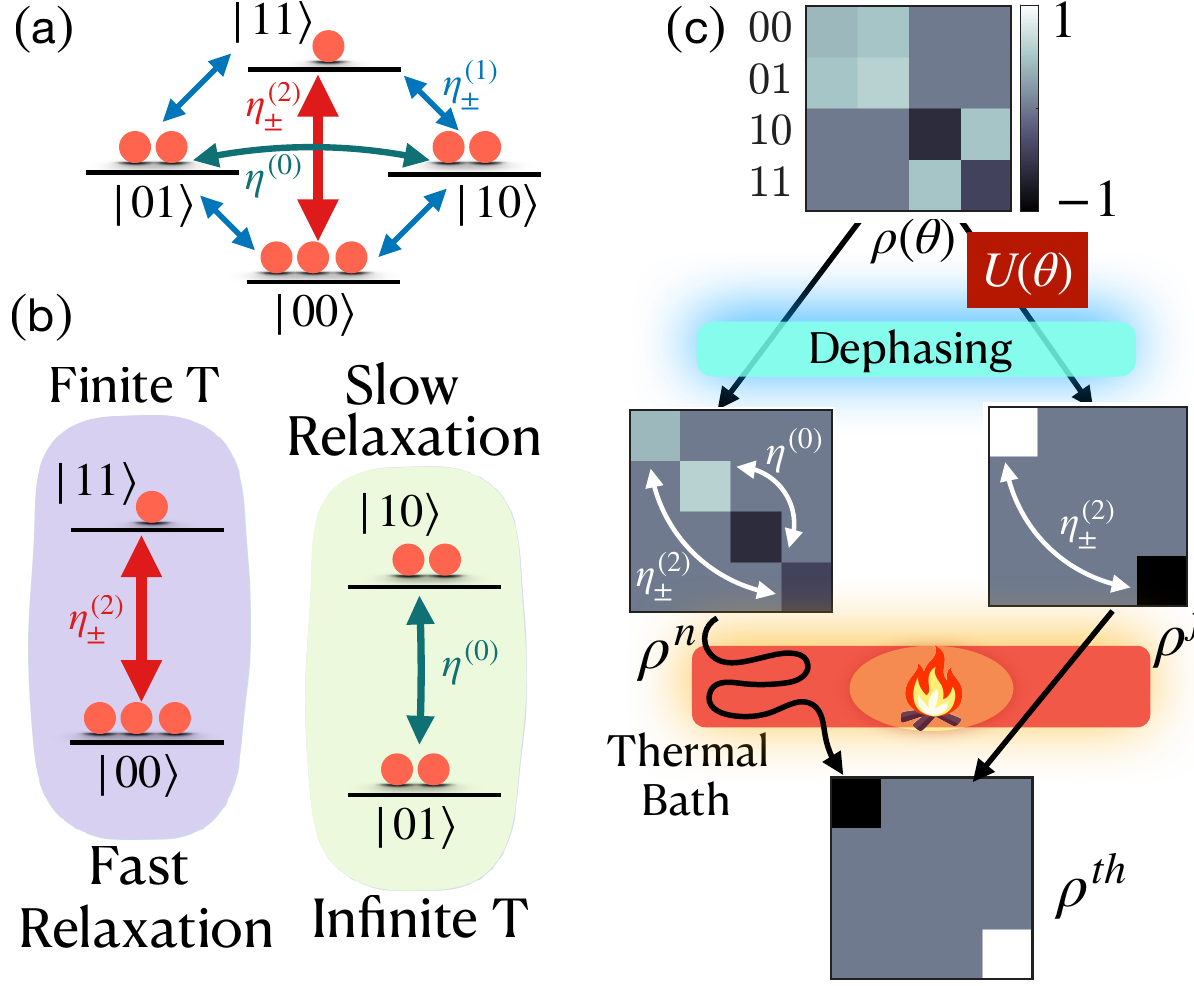} 
\caption{(a) The energy eigenstates of the spin Hamiltonian $\mathcal{H}(t)$ (considering $\omega_0$ to be negative), with various transitions induced by $\mathcal{L}_p$. (b) For particular choices of initial states, the dynamics under $\mathcal{L}_p$ can be interpreted as thermalization of two independent qubits. (c) A schematic view of the experiments that observe quantum Mpemba effect.} 
\label{fig:thseq}
\end{figure}

At infinite temperature ($\epsilon=0$), it can be shown (see Appendix A) that if the initial state $\rho(0)$ contains only population terms satisfying 
\begin{equation}
p_{00}+p_{11} = p_{01} + p_{10},
\label{eq:incond}
\end{equation}
then the coherence terms remain zero in the subsequent dynamics, where $p_{ab}$ is the coefficient of the $\proj{ab}$ term in $\rho(0)$. However, since we are working at high but finite temperature,  we have to consider up to first order of $\epsilon$, as in Eq.~\Eqref{eq:thstate}. Interestingly, if we start with only population terms respecting Eq.~\Eqref{eq:incond}, the coherence terms generated in the evolution will be of the order of $(K_0/\Delta)\epsilon$, which is a very small number and can thus be neglected as long as $K_0 \ll \Delta$. For such initial states, which can be conveniently represented as a vector $\vec{p}(0):=(p_{00},p_{01},p_{10},p_{11})$, the time evolution can be described as (see Appendix A for details) 
\begin{equation}
\partial_t \vec{p}(t)=\mathcal{L}_p\, \vec{p}(t) \implies \vec{p}(t) = \vec{p}_{th} +  \sum_{n=1}^{3}a_n\, \expo{\lambda_nt}\, \vec{v}_n, \label{eq:restrict}
\end{equation}
where the overlaps $a_n=\vec{w}_n\cdot \vec{p}(0)$, the population at thermal state is $\vec{p}_{th}=(1+2\epsilon,1,1,1-2\epsilon)/4$, and the decay rates are $\{\lambda_1,\lambda_2,\lambda_3\}=-(K_0/24)\{5,6,15\}$, with $\vec{v}_n(\vec{w}_n)$ being the right (left) eigenvector of $\mathcal{L}_p$ w.r.t decay rate $\lambda_n$. The transitions between the population terms $\proj{ab}\rightarrow\proj{cd}$ for $a,b,c,d \in \{0,1\}$ induced by $\mathcal{L}_p$ can be divided into three categories depending on $m=(c+d)-(a+b)$ (see Fig.~\ref{fig:thseq}~(a)). For $m=\pm1$, we get single quantum transitions with decay rates $\eta_{\pm}^{(1)}=K_0(1\mp \epsilon)/16$, for $m=\pm2$ we get double quantum transitions with decay rates $\eta_{\pm}^{(2)}=K_0(1\mp 2\epsilon)/4$, and for $m=0$ we get zero quantum transitions with decay rate $\eta^{(0)}=K_0/24$. Note that the transition probabilities obey quantum detailed balance, where temperature corrections are considered up to the first order in $\epsilon$. 

Interestingly, if we choose initial states such that the dynamics happens only through $\eta_{\pm}^{(2)}$ and $\eta^{(0)}$, then the thermalization process of two spins can be described as independent thermalization of two qubits (see Fig.~\ref{fig:thseq}~(b)), with the first one (composed of states $\Ket{00}$ and $\Ket{11}$) is in contact with a bath at temperature $T$ while the second one (composed of states $\Ket{01}$ and $\Ket{10}$) remains in contact with an infinite temperature bath. Since $\eta_{\pm}^{(2)} > \eta^{(0)}$, a state which can go to thermal equilibrium only through the double quantum transitions will relax much faster than a state which requires single or zero quantum transitions to thermalize. This observation will help us to find suitable nuclear spin states which can show quantum Mpemba effect.       

\emph{Observing Quantum Mpemba Effect--} To experimentally investigate quantum Mpemba effect, we consider the two spin-1/2 $^{1}$H nuclei of  $2$-Chloroacrylonitrile (CAN), dissolved in Dimethyl Sulfoxide (DMSO) ($20~\mu$L of CAN in $600~\mu$L of DMSO), as our two spin system, which sits in a Zeeman field of $B_0 = 11.7$ T at an ambient temperature of $T=295$ K inside a $500$ MHz Bruker NMR spectrometer. The spin Hamiltonian and the equilibrium thermal state of the system is described by Eqs.~\ref{eq:nmr_Ham} and \ref{eq:thstate}, respectively, with $\Delta/2\pi= 89$ Hz, $J=3.24$ Hz, and $\omega_0/2\pi = 500.02$ MHz. 
We consider the state 
\begin{gather}
\rho(\theta)= \mathbb{1}/4 + 2\epsilon[\cos^2 \theta (I_{1z}+I_{2z})+ \nonumber \\
\sin^2{\theta}(I_{1z}-I_{2z})+  \sin(2\theta)I_{2y}]/4,~\text{for}~\theta \in (0,\pi/2), \label{eq:intern_state}
\end{gather}
to study the relaxation dynamics. The experimental procedure is sketched in Fig.~\ref{fig:thseq}~(c)(see Appendix C for details). After preparation, the state of Eq.~\Eqref{eq:intern_state} is subjected to a strong inhomogeneous magnetic field using a pulsed field gradient (PFG), which eliminates the terms of the density matrix that are susceptible to inhomogeneous dephasing and ensures the subsequent evolution to happen solely under natural thermalization as described by Eq.~\Eqref{eq:GSKL}. The state of Eq.~\Eqref{eq:intern_state} gets dephased by the PFG and becomes   $\rho^{n}(\theta) = \mathbb{1}/4 + 2\epsilon (I_{1z} + I_{2z} \cos{2\theta})/4$, which is then allowed to thermalize naturally. Note that $\rho^{n}(\theta)$ satisfies Eq.~\Eqref{eq:incond}, and hence it evolves under $\mathcal{L}_p$ according to Eq.~\Eqref{eq:restrict}.  As shown in Fig.~\ref{fig:thseq}~(c), it requires the fast ($\eta_{\pm}^{(2)}$) as well as the slow ($\eta^{(0)}$) transition for $\rho^{n}(\theta)$ to reach the thermal equilibrium. The trace distance from $\rho^{n}(\theta)$ to the equilibrium thermal state $\rho^{th}$ reads $D(\rho^{n}(\theta),\rho^{th}) = \epsilon |1-\cos{2\,\theta}|$. However, by applying an unitary operator $U(\theta)$ on $\rho(\theta)$, we  push it further away from equilibrium, which passes through the dephasor and becomes  $\rho^{f} = \mathbb{1}/4 - \epsilon(I_{1z}+I_{2z})/2$  such that $D(\rho^{f},\rho^{th}) = \epsilon > D(\rho^{n}(\theta),\rho^{th})$. The transformed state $\rho^f$ could thermalize only via the fast $\eta_{\pm}^{(2)}$ transition. In other words, $\rho^{f}$ possesses zero overlap with the slowest decay mode $\lambda_1$ of $\mathcal{L}_p$, and thus, despite of being much further away, should reach equilibrium in shorter time than $\rho^{n}$.       

\begin{figure}
\includegraphics[width=8.6cm, clip=true, trim={0.7cm 2.1cm 13.7cm 2.4cm}]{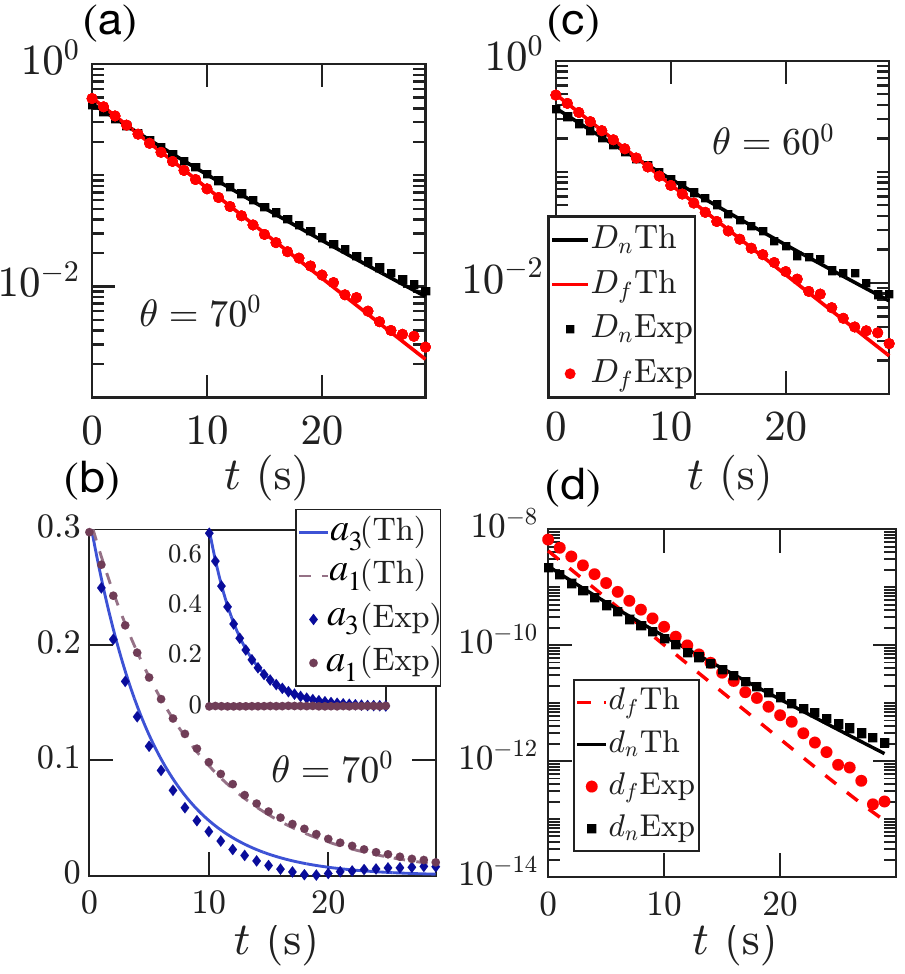} 
\caption{(a,c) Occurrence of Quantum Mpemba Effect where the far state $\rho^f$ relaxes much faster than the nearer state $\rho^n$ for different values of $\theta$. As time elapses, the distance of $\rho^f$ and $\rho^n$ from thermal state $\rho^{th}$ is measured by the trace distance $D_n$ and $D_f$, respectively. (b) $\rho^n$ overlaps with both decay modes $\lambda_1$ and $\lambda_3$ while (inset) $\rho^f$ overlaps only with the fast decay mode $\lambda_3$, the data(s) are for experiment at $\theta=70^0$. The y axis is scaled by multiplying with $1/(2\epsilon)$ in (a-c). (d) Experimental demonstration of `genuine quantum Mpemba effect' where an initial state $\rho^f$ with higher relative entropy (w.r.t $\rho^{th}$), as measured by $d_f$ attains equilibrium faster than a state $\rho^n$ with lower relative entropy, as measured by $d_n$. The experimental errors lie well within the marker size used. The theoretical predictions are made at $\tau_c=2.1$ ps and $b=5.903$ KHz.} 
\label{fig:exp}
\end{figure}

The experimental results are shown in Fig~\ref{fig:exp} (a) and (c), which confirms the theoretical prediction and demonstrates the quantum Mpemba effect in natural thermalization of nuclear spins. We also experimentally measure the overlap of each state $\rho^f$ and $\rho^n$ with the respective decay modes. As shown in Fig.~\ref{fig:exp}~(b), the state $\rho^n$ overlaps with both fast ($\lambda_3$) and slow ($\lambda_1$) decay modes and hence thermalises slowly, while $\rho^f$ only overlaps with the fast decay mode  (Fig.~\ref{fig:exp}~(b) inset) and hence goes to thermal equilibrium much faster. Thus we experimentally verify the cause behind the existence of observable quantum Mpemba effect in nuclear spins to be the dominance of $\eta_{\pm}^{(2)}$ transitions over $\eta^{(0)}$ transition under dipolar relaxation, as predicted by the theory. After the preparation of $\rho^f$ and $\rho^n$, the nuclear spins are left to evolve without applying any pulse up to $29$ seconds, which extends well beyond their $T_1 = 5.7$ s to ensure the observed Mpemba comes from their natural thermalization process without any bath engineering. Within the limitations of experiments on an open quantum system, we observe remarkable agreement between theoretical prediction and experimental data.

\emph{Observing Genuine Mpemba Effect}--- It has been proposed recently \cite{PhysRevLett.133.140404} that in order to observe `genuine quantum Mpemba effect' from a thermodynamic perspective, one should use the non equilibrium free energy, which equals the relative entropy $d(\rho_1,\rho_2)=-\text{tr}[\rho_1(\log \rho_1 - \log \rho_2)]$ up to an additive constant, as a metric.  We identify the favorable states for observing the genuine  Mpemba effect to be
$\rho^n = \mathbb{1}/4 + \epsilon(I_{1z}-I_{2z})/2$ and $\rho^f = \mathbb{1}/4 - \epsilon(I_{1z}+I_{2z})/2$.  We first prepare the state $\rho^n$, which overlaps with both decay modes $\lambda_1$ and $\lambda_3$. As shown in Fig.~\ref{fig:thseq}~(c), after passing it through a strong dephasor, we let the state to thermalize naturally while measuring $d(\rho^n,\rho^{th})$ with time. The state $\rho^n$ is converted to $\rho^f$ by applying a $\pi$-pulse on the first spin, which overlaps only with the fastest decay mode $\lambda_3$. Even though $\rho^f$ contains higher free energy than $\rho^n$, as shown in Fig~\ref{fig:exp}~(d), under natural thermalization it attains the free energy value of the equilibrium state much faster than $\rho^n$. Thus we  experimentally demonstrate the existence of genuine quantum Mpemba effect in the natural thermalization of nuclear spins.

\emph{Summary and Outlook}--- We have investigated the quantum Mpemba effect in the natural thermalization of nuclear spins. By identifying dipolar relaxation as the dominant decoherence process, we derived the conditions under which the effect can emerge. Preparing nuclear spin states that satisfy these conditions, we observed that the state farther from equilibrium relaxes faster than the one closer to equilibrium, thereby establishing the quantum Mpemba effect as an intrinsic feature of natural thermalization, requiring no external bath engineering. In addition, we have provided the experimental demonstration of the genuine quantum Mpemba effect in a natural relaxation process.

As quantum information technologies continue to advance, mechanisms that shorten relaxation times are of critical importance, as they directly reduce idle periods between consecutive runs of quantum devices. Our experiments highlight that the Mpemba effect,long debated in the classical domain, arises naturally in quantum spin systems, which is the hardware of many present day quantum devices. Beyond resolving a fundamental question in nonequilibrium statistical mechanics, these findings open pathways toward exploiting the Mpemba effect as a resource for quantum technologies, where accelerated equilibration could be harnessed for more efficient initialization, error mitigation, and control of large-scale quantum devices.



\emph{Acknowledgments}--- AC acknowledges Infosys Foundation Travel award and the organisers of QTD 2025, Singapore. TSM acknowledges funding from I-HUB QTF.

\bibliography{bibliography}

\begin{center}
\textbf{Appendix}
\end{center}
\subsection{A : Dynamics under the Dipolar Relaxation}
The expressions of the spherical tensor operators $\{T_{2m}\}_{m=-2}^{2}$ mentioned in Eq.~\Eqref{eq:GSKL} of main text are given below 
\begin{gather}
T_0 = \frac{1}{\sqrt{6}} \left(3I_{1z}\,I_{2z} - \vec{I_1} \cdot \vec{I_2} \right) = T_{0}^{\dagger}, \nonumber \\
T_{\pm1} = \mp \frac{1}{2}
\left(I^{1}_{\pm}I_{2z} + I_{1z}I^{2}_{\pm}  \right) = -T^{\dagger}_{\mp1} \nonumber \\
T_{\pm2}= \frac{1}{2} I^{1}_{\pm}\,I^{2}_{\pm} = T^{\dagger}_{\pm2} \label{eq:spher_tens},    
\end{gather}
where $\vec{I_{n}} = (I_{nx},I_{ny},I_{nz})$ ,  and $I_{\pm}^{n}=I_{nx} \pm I_{ny}$ for the $n$'th spin. 


In the ZQB, the coefficients of the population terms $\proj{ab}$ are $p_{ab}$, and the coefficient of the coherence term $\outpr{01}{10}$ is $c$. The GSKL equation can be written now in the ZQB block as  
\begin{align}
    \partial_t 
\begin{pmatrix}
p_{00} \\
p_{01} \\
p_{10} \\
p_{11} \\
c \\
c^*
\end{pmatrix}=\mathcal{L}_0
\begin{pmatrix}
p_{00} \\
p_{01} \\
p_{10} \\
p_{11} \\
c \\
c^*
\end{pmatrix}
\end{align}
with 
\begin{align}
\mathcal{L}_0 =
\begin{pmatrix}
A & B \\
C & D
\end{pmatrix},
\end{align}
where
\begin{widetext}
\begin{equation}
A =
\begin{pmatrix}
-\tfrac{1}{8} K_0 (1 - \epsilon) - \tfrac{1}{4} K_0 (1 - 2 \epsilon) & 
\tfrac{1}{16} K_0 (1 + \epsilon) & 
\tfrac{1}{16} K_0 (1 + \epsilon) \\
\tfrac{1}{16} K_0 (1 - \epsilon) & 
-\tfrac{K_0}{24} - \tfrac{1}{16} K_0 (1 + \epsilon) - \tfrac{1}{16} K_0 (1 - \epsilon) & 
\tfrac{K_0}{24} \\
\tfrac{1}{16} K_0 (1 - \epsilon) & 
\tfrac{K_0}{24} & 
-\tfrac{K_0}{24} - \tfrac{1}{16} K_0 (1 + \epsilon) - \tfrac{1}{16} K_0 (1 - \epsilon)
\end{pmatrix},
\end{equation}
\begin{equation}
B = 
\begin{pmatrix}
\tfrac{1}{4} K_0 (1 + 2 \epsilon) & 
\tfrac{1}{16} K_0 (1 + \epsilon) & 
\tfrac{1}{16} K_0 (1 + \epsilon) \\
\tfrac{1}{16} K_0 (1 + \epsilon) & 
-\tfrac{1}{32} K_0 (1 + \epsilon) - \tfrac{1}{32} K_0 (1 - \epsilon) & 
-\tfrac{1}{32} K_0 (1 + \epsilon) - \tfrac{1}{32} K_0 (1 - \epsilon) \\
\tfrac{1}{16} K_0 (1 + \epsilon) & 
-\tfrac{1}{32} K_0 (1 + \epsilon) - \tfrac{1}{32} K_0 (1 - \epsilon) & 
-\tfrac{1}{32} K_0 (1 + \epsilon) - \tfrac{1}{32} K_0 (1 - \epsilon)
\end{pmatrix},
\end{equation}
\begin{equation}
C =
\begin{pmatrix}
\tfrac{1}{4} K_0 (1 - 2 \epsilon) & 
\tfrac{1}{16} K_0 (1 - \epsilon) & 
\tfrac{1}{16} K_0 (1 - \epsilon) \\
\tfrac{1}{16} K_0 (1 - \epsilon) & 
-\tfrac{1}{32} K_0 (1 + \epsilon) - \tfrac{1}{32} K_0 (1 - \epsilon) & 
-\tfrac{1}{32} K_0 (1 + \epsilon) - \tfrac{1}{32} K_0 (1 - \epsilon) \\
\tfrac{1}{16} K_0 (1 - \epsilon) & 
-\tfrac{1}{32} K_0 (1 + \epsilon) - \tfrac{1}{32} K_0 (1 - \epsilon) & 
-\tfrac{1}{32} K_0 (1 + \epsilon) - \tfrac{1}{32} K_0 (1 - \epsilon)
\end{pmatrix},
\end{equation}
\begin{equation}
D =
\begin{pmatrix}
-\tfrac{1}{4} K_0 (1 + 2 \epsilon) - \tfrac{1}{8} K_0 (1 + \epsilon) & 
\tfrac{1}{16} K_0 (1 - \epsilon) & 
\tfrac{1}{16} K_0 (1 - \epsilon) \\
\tfrac{1}{16} K_0 (1 + \epsilon) & 
i \Delta - \tfrac{K_0}{24} - \tfrac{1}{16} K_0 (1 + \epsilon) - \tfrac{1}{16} K_0 (1 - \epsilon) & 
\tfrac{K_0}{24} \\
\tfrac{1}{16} K_0 (1 + \epsilon) & 
\tfrac{K_0}{24} & 
- i \Delta - \tfrac{K_0}{24} - \tfrac{1}{16} K_0 (1 + \epsilon) - \tfrac{1}{16} K_0 (1 - \epsilon)
\end{pmatrix}.
\end{equation}
\end{widetext}
Now we can easily show that at infinite temperature ($\epsilon=0$), the population and coherence satisfy the following equation 
\begin{align}
\frac{d}{dt}
\begin{pmatrix}
X_{1}(t) \\
X_{2}(t) \\
X_{3}(t)
\end{pmatrix}
=
\begin{pmatrix}
-\tfrac{K_{0}}{4} & \tfrac{K_{0}}{4} & 0 \\
\tfrac{K_{0}}{8} & -\tfrac{K_{0}}{8} & i \Delta \\
0 & i \Delta & -\tfrac{5K_{0}}{24}
\end{pmatrix}
\begin{pmatrix}
X_{1}(t) \\
X_{2}(t) \\
X_{3}(t)
\end{pmatrix}\;,
\end{align}
where $ X_{1}=p_{00}+p_{11}-(p_{01}+p_{10}), X_{2}=c+c^*$ and $X_{3}=c-c^*$.
If the initial density matrix contains only population terms satisfying Eq.~\Eqref{eq:incond} of the main text i.e., $X_{1}(0)=X_{2}(0)=X_{3}(0)=0$, then the coherence terms remain zero throughout the subsequent dynamics. This further reduces the space from $6 \times 6$  to $4 \times 4$. However, since we are working at high but finite temperature, we have to consider up to first order of $\epsilon$. Considering that we can explicitly show that the order of coherence that the system generates will be of the order $ (K_0/\Delta) \epsilon$. This is very a small number as long as $K_0 \ll \Delta$, and so we can neglect this. Therefore, we can safely say even at finite but high temperatures, for initial states containing only population terms satisfying Eq.~\Eqref{eq:incond}, the population and dynamics remain decoupled from the coherence terms :  
\begin{align}
\partial_t 
\begin{pmatrix}
p_{00} \\
p_{01} \\
p_{10} \\
p_{11} 
\end{pmatrix}=\mathcal{L}_p
\begin{pmatrix}
p_{00} \\
p_{01} \\
p_{10} \\
p_{11} 
\end{pmatrix}
\end{align}
with
\begin{widetext}
\begin{equation}
\mathcal{L}_p =
\end{equation}
\begin{equation}
\scalebox{0.85}{$
\begin{bmatrix}
-\tfrac{1}{8} K_0 (1 - \epsilon) - \tfrac{1}{4} K_0 (1 - 2 \epsilon) & 
\tfrac{1}{16} K_0 (1 + \epsilon) & 
\tfrac{1}{16} K_0 (1 + \epsilon) & 
\tfrac{1}{4} K_0 (1 + 2 \epsilon) \\
\tfrac{1}{16} K_0 (1 - \epsilon) & 
-\tfrac{K_0}{24} - \tfrac{1}{16} K_0 (1 + \epsilon) - \tfrac{1}{16} K_0 (1 - \epsilon) & 
\tfrac{K_0}{24} & 
\tfrac{1}{16} K_0 (1 + \epsilon) \\
\tfrac{1}{16} K_0 (1 - \epsilon) & 
\tfrac{K_0}{24} & 
-\tfrac{K_0}{24} - \tfrac{1}{16} K_0 (1 + \epsilon) - \tfrac{1}{16} K_0 (1 - \epsilon) & 
\tfrac{1}{16} K_0 (1 + \epsilon) \\
\tfrac{1}{4} K_0 (1 - 2 \epsilon) & 
\tfrac{1}{16} K_0 (1 - \epsilon) & 
\tfrac{1}{16} K_0 (1 - \epsilon) & 
-\tfrac{1}{4} K_0 (1 + 2 \epsilon) - \tfrac{1}{8} K_0 (1 + \epsilon)
\end{bmatrix}
$}
\end{equation}
\end{widetext}
We can find the eigenvalues and right eigenvectors of $\mathcal{L}_p$, which are given by
\begin{align*}
v_{0}&= \vec{p}_{th} = \frac{1}{4}
(\, 1 + 2\epsilon,\, 1,\, 1,\, 1 - 2\epsilon \,),~ \lambda_{0}=0;\\
v_{1} &= 
\frac{1}{\sqrt{2}} 
(\, 0,\; -1,\; 1,\; 0 \,),  ~\lambda_{1}=-\frac{5 K_0}{24};\\
v_{2} &= 
\frac{1}{\sqrt{\,4 + \tfrac{8}{3}\epsilon(5 + 4\epsilon)}} 
(\, 1 + \tfrac{16}{3}\epsilon,\; -1 - \tfrac{8}{3}\epsilon,\; -1 - \tfrac{8}{3}\epsilon,\; 1 \,),  \\ &\lambda_{2}=-\frac{K_0}{4};\\
v_{3} &= 
\frac{1}{\sqrt{\tfrac{2}{3}(1-\epsilon)(3-5\epsilon)}} 
(\, -1 + \tfrac{2}{3}\epsilon,\; -\tfrac{\epsilon}{3},\; -\tfrac{\epsilon}{3},\; 1 \,), \\ &\lambda_{3}=-\frac{5 K_0}{8}.
\end{align*}
Similarly, the left eigenvectors of $\mathcal{L}_p$ are given by
\begin{align*}
w_{0} &= 
(\, 1,\; 1,\; 1,\; 1 \,), \\
w_{1} &= 
\frac{1}{\sqrt{2}} 
(\, 0,\; -1,\; 1,\; 0 \,), \\
w_{2} &= 
\frac{1}{\sqrt{\,4 + \tfrac{8}{3}\epsilon(5 + 4\epsilon)}} 
(\, 1 + \tfrac{4}{3}\epsilon,\; -1 - \tfrac{2}{3}\epsilon,\; -1 - \tfrac{2}{3}\epsilon,\; 1 \,), \\
w_{3} &= 
\frac{1}{\sqrt{\tfrac{2}{3}(1-\epsilon)(3-5\epsilon)}} 
(\, -1 + \tfrac{14}{3}\epsilon,\; -\tfrac{\epsilon}{3},\; -\tfrac{\epsilon}{3},\; 1 \,).
\end{align*}
In all the above expressions, terms upto first order of $\epsilon$ is to be considered.
If we now take the initial state to be 
\begin{align}
\{& \tfrac{1}{4} + (1+\cos{2 \theta})\epsilon/2,\; 
\tfrac{1}{4} + (1-\cos{2 \theta})\epsilon/2 \nonumber\\
&\tfrac{1}{4} + (-1+\cos{2 \theta})\epsilon/2,\; 
\tfrac{1}{4} + (-1-\cos{2 \theta})\epsilon/2 \}
\end{align}
then we can show that the population of the density matrix at any time are given by
\[
\begin{aligned}
\rho_{11}(t) &= \tfrac{1}{4} + \Big( -\tfrac{1}{2} + \frac{(2+ \cos{2 \theta})}{2}\, e^{-\tfrac{5 g t}{8}} \Big)\, \epsilon, \\[6pt]
\rho_{44}(t) &= \tfrac{1}{4} + \Big( \tfrac{1}{2} - \frac{(2+ \cos{2 \theta})}{2}\, e^{-\tfrac{5 g t}{8}} \Big)\, \epsilon, \\[6pt]
\rho_{22}(t) &= \tfrac{1}{4} - \frac{1}{2}\big( -1+\cos{2 \theta} \big)\, e^{-\tfrac{5 g t}{24}}\, \epsilon, \\[6pt]
\rho_{33}(t) &= \tfrac{1}{4} + \frac{1}{2}\big( -1+\cos{2 \theta} \big)\, e^{-\tfrac{5 g t}{24}}\, \epsilon.
\end{aligned}
\]

Now $\rho_{11}(t), \rho_{44}(t)$ are decoupled from the other two components and independently satisfy the trace condition. This means the dynamics can be interpreted as two non-interacting qubits, each dissipating through its own bath. Interestingly, while one qubit perceives the bath as being at the same temperature as the original environment, the other effectively experiences the bath as being at infinite temperature.

In summary, the complicated two-qubit dynamics can be mapped to a simple effective picture of two non-interacting qubits, each dissipating through its own bath at a different temperature.

\subsection{B : The Master Equation from a Semi-Classical Microscopic Picture} \label{appen:GSKL}
We consider a sample whose molecule contains two spin $1/2$ nuclei of same species (two homo-nuclear qubits). The sample is dissolved in some non reacting solvent to prepare a dilute solution (approximately $\approx 50~\mu$L sample in $\approx 600~\mu$L solvent), and kept inside an NMR spectrometer with s a strong Zeeman field $B_0\,\hat{z}$. The Hamiltonian, as seen by the spin system of a single molecule in the ensemble, reads
\begin{gather}
\mathcal{H}(t) = \mathcal{H}_0 + \mathcal{H}_{\text{fluc}}(t), \nonumber \\
\text{where}~\mathcal{H}_{0} = \underbrace{\omega_0\,\left(I_{1z} + I_{2z}\right)}_{\mathcal{H}_A} - \underbrace{\delta\,\left(I_{1z} + I_{2z}\right) + 2\pi J\, \vec{I_1}\cdot \vec{I_2}}_{\mathcal{H}_B} 
\nonumber
\end{gather}
is the Hamiltonian mentioned in Eq.~\Eqref{eq:nmr_Ham} and it does not depend on the molecular orientation. Hence, it remains invariant in time while the sample molecule undergo rapid molecular tumbling due to diffusion in liquids at finite temperature. Moreover, at a given instant of time $t$, $\mathcal{H}_0$ is same for all the molecules of the ensemble distributed in random orientations, and hence it is called the coherent part of the Hamiltonian. Whereas $\mathcal{H}_{\text{fluc}}(t)$ describes the interactions that depends on molecular orientation. It varies randomly with time since the molecular orientation undergoes rapid tumbling due to the diffusional motion of the sample molecule in the liquid solvent at finite temperature. We fix a frame of reference, called the Lab frame $(x,y,z)$,  which stays fixed in time having $z$ axis coinciding with the Zeeman field. To take care of the random molecular tumbling, we consider the frame of reference formed by the principal axis system (PAS) of the chemical shift anisotropy tensor $(X,Y,Z)$ \footnote{assuming same PAS for both the nuclei} \cite{goldman2001quantum,kowalewski2017nuclear}, which will be called the molecular frame as it remains glued with the molecule and undergoes random rotation in time with respect to the lab frame $(x,y,z)$. For simplicity, we also consider that the vector connecting both the nuclei $\hat{n}$, which also remains glued with the molecule and undergoes random rotation with respect to the lab frame, coincides with the $Z$ axis of the molecular frame \footnote{For a more general treatment, consider \cite{KUMAR2000191}}. The fluctuating Hamiltonian contains two interactions  : the intra-molecular dipole-dipole interaction $\mathcal{H}_{\text{DD}}(t)$ and the chemical shift anisotropy $\mathcal{H}_{\text{CSA}}(t)$ (assumed to be axially symmetric). Utilizing the freedom to choose the initial time, we consider that the molecular frame coincides with the Lab frame at $t=0$. At a later time $t$, the fluctuating Hamiltonian $\mathcal{H}_{\text{fluc}}(t)$ can be written in the Lab frame as   \cite{goldman2001quantum,GOLDMAN2001160,kowalewski2017nuclear}
\begin{gather} 
\mathcal{H}_{\text{fluc}}(t) = \mathcal{H}_{\text{DD}}(t) + \mathcal{H}_{\text{CSA}} (t),~~\text{where} \nonumber \\
\mathcal{H}_{\text{DD}}(t) = -\underbrace{\frac{\mu_0}{4\pi}\frac{\gamma^2\,\hbar}{r^3}}_{b} \sum_{m=-2}^{2} A^{*}_{2m}\left(\theta(t),\phi(t) \right)\,T_{2m}(\vec{I_1},\vec{I_2}), \nonumber \\
\mathcal{H}_{\text{CSA}}^{1,2}(t) = \underbrace{\frac{\gamma\,\Delta}{3}}_{d} \sum_{m=-1}^{1} A^{*}_{2m}\left(\theta(t),\phi(t) \right)\,T_{2m}(\vec{B},\vec{I}_{1,2}), \nonumber  
\end{gather}
with $r$ being the distance between the two nuclei $(\approx 1-4 \, A^0)$, $\Delta$ being the chemical shift anisotropy \cite{goldman2001quantum}, and both the spin vectors $\vec{I_1}$,$\vec{I_2}$ and the Zeeman field vector $\vec{B}$ is written in the Lab frame. The angles $\theta(t)$ and $\phi(t)$ are the co-latitude and longitude, respectively, of the molecular frame $\hat{Z}$ axis with respect to the Lab frame and are random functions of time. The exact forms of the spherical functions are
\begin{gather}
A_{0}(\theta,\phi) = \sqrt{\frac{3}{2}} \left(3\cos^2 \theta -1 \right) = A_{0}^*(\theta,\phi), \nonumber \\
A_{\pm1}(\theta,\phi)  =\mp \,3 \,\sin \theta \cos \theta \expo{\pm i\phi} = \mp A^{*}_{\mp 1}(\theta,\phi), \nonumber \\
A_{\pm 2}(\theta,\phi) = \frac{3}{2} \,\sin^2 \theta \expo{\pm 2i\phi}
= A^{*}_{\mp 2}. \label{eq:spher_har}
\end{gather}
Whereas the spherical tensor operators $T_{2m}(\vec{A},\vec{B})$ reads 
\begin{gather}
T_0(\vec{A},\vec{B}) = \frac{1}{\sqrt{6}} \left(3A_{z}\,B_{z} - \vec{A} \cdot \vec{B} \right) = T_{0}^{\dagger}(\vec{A},\vec{B}), \nonumber \\
T_{\pm1}(\vec{A},\vec{B}) = \mp \frac{1}{2}
\left(A_{\pm}B_{z} + A_{z}B_{\pm}  \right) = -T^{\dagger}_{\mp1}(\vec{A},\vec{B}) \nonumber \\
T_{\pm2}(\vec{A},\vec{B}) = \frac{1}{2} A_{\pm}\,B_{\pm} = T^{\dagger}_{\pm2}(\vec{A},\vec{B}) \label{eq:spher_tens}, 
\end{gather}
with $A_{\pm} := A_x \pm i A_{y}$ and same for $B_{\pm}$. Let $\varrho$ be the density operator of the spin system of one particular molecule. It, and the Hamiltonian $\mathcal{H}(t)$ transforms as we go to the interaction frame defined by $U_0 = \exp(i\mathcal{H}_A t)$ as :
\begin{gather}
\mathcal{H}(t) \rightarrow \widetilde{\mathcal{H}}(t) = U_0(t)\mathcal{H}(t)\,U_{0}^{\dagger}(t) - i U_0(t)U_{0}^{\dagger}(t) \nonumber \\ = \mathcal{H}_{B} + \underbrace{U_0(t)\mathcal{H}_{\text{fluc}}(t)U^{\dagger}_0(t)}_{\widetilde{\mathcal{H}}_{\text{fluc}}(t)}, \nonumber \\
\varrho(t) \rightarrow \widetilde{\varrho}(t) = U_0(t)\varrho(t)\,U_{0}^{\dagger}(t). 
\end{gather}
The equation of motion in the interaction frame can be readily written as \cite{sakurai1986modern}
\begin{gather}
\frac{d\widetilde{\varrho}(t)}{dt} = -i[\mathcal{H_B}+\widetilde{\mathcal{H}}_{\text{fluc}}(t),\widetilde{\varrho}(t)]. \label{eq:VN}
\end{gather}
The time by which $\widetilde{\varrho}(t)$ changes is typically determined by the norm of the generators of the motion, i.e $\dnorm{\mathcal{H}_B}$ and $\dnorm{\widetilde{\mathcal{H}}_{\text{fluc}}(t)}$. We assume this time to be much much larger than the typical time $(\tau_c)$ by which the rapidly fluctuating Hamiltonian $\widetilde{\mathcal{H}}_{\text{fluc}}(t)$ changes. Therefore, it is possible to consider a time $t$ \cite{GOLDMAN2001160} such that 
\begin{gather}
\tau_c \ll t \ll \frac{1}{\dnorm{\mathcal{H}_B}},\frac{1}{\dnorm{\widetilde{\mathcal{H}}_{\text{fluc}}(t)}}, \label{eq:COND}
\end{gather}
where $\widetilde{\varrho}(t) \approx \varrho(0)$. In such a time $t$, the formal solution  can be written  as
\begin{gather}
\widetilde{\varrho}(t) = \varrho(0) - i\int_{0}^{t}dt_1 [\mathcal{H_B}+\widetilde{\mathcal{H}}_{\text{fluc}}(t_1),\widetilde{\varrho}(t_1)].
\end{gather}
Putting this back to Eq.~\ref{eq:VN} \cite{breuer2002theory,lidar2020lecturenotestheoryopen} we get
\begin{gather}
\frac{d\widetilde{\varrho}(t)}{dt} = -i[\mathcal{H_B}+\widetilde{\mathcal{H}}_{\text{fluc}}(t),\varrho(0)] \nonumber \\ - \int_{0}^{t} dt_1 [\mathcal{H_B}+\widetilde{\mathcal{H}}_{\text{fluc}}(t), [\mathcal{H_B}+\widetilde{\mathcal{H}}_{\text{fluc}}(t_1),\varrho(0)]] \nonumber \\
= -i[\mathcal{H_B}+\widetilde{\mathcal{H}}_{\text{fluc}}(t),\widetilde{\varrho}(t)] \nonumber \\ - \int_{0}^{t} dt_1 [\mathcal{H_B}+\widetilde{\mathcal{H}}_{\text{fluc}}(t), [\mathcal{H_B}+\widetilde{\mathcal{H}}_{\text{fluc}}(t_1),\widetilde{\varrho}(t)]].
\label{eq:apr1}
\end{gather}
We note that in the experiment we do not have any access to this density operator $\widetilde{\varrho}(t)$ belonging to one particular molecule. Therefore we define the density operator $\widetilde{\rho}(t)=\overline{\widetilde{\varrho}(t)}$ describing the spin system's state of the ensemble of all the molecules in the tube, where $\overline{(\cdot)}$ dictates ensemble average. We assume that the density operator $\widetilde{\varrho}(t)$ and the $\widetilde{\mathcal{H}}_{\text{fluc}}(t)$ are statistically uncorrelated so that they can be averaged separately \cite{abragam1961principles,cavanagh1996protein}. After averaging, Eq.~\Eqref{eq:apr1} reads
\begin{gather}
\frac{d\widetilde{\rho}(t)}{dt} = \frac{d\overline{\widetilde{\varrho}(t)}}{dt} = -i[\mathcal{H_B},\widetilde{\rho}(t)] \nonumber \\
- \int_{0}^{t}dt_1 [\overline{\widetilde{\mathcal{H}}_{\text{fluc}}(t), [\widetilde{\mathcal{H}}_{\text{fluc}}(t_1)},\widetilde{\rho}(t)]],
\end{gather}
where we have kept only up to the first non-zero terms \cite{GOLDMAN2001160} of $\widetilde{\mathcal{H}}_B$ and $\widetilde{\mathcal{H}}_{\text{fluc}}(t)$, after noticing that $\overline{\widetilde{\mathcal{H}}_{\text{fluc}}(t)}=0$ since the spherical functions of Eq.~\Eqref{eq:spher_har} averages out to zero over an uniform distribution of $\theta(t)$ and $\phi(t)$. We assume that the fluctuating components of the Hamiltonian can be represented by a weakly stationary process such that after a change of variable $\tau = t-t_1$ :
\begin{gather}
- \int_{0}^{t}d\tau [\overline{\widetilde{\mathcal{H}}_{\text{fluc}}(t), [\widetilde{\mathcal{H}}_{\text{fluc}}(t-\tau)},\widetilde{\rho}(t)]] \nonumber \\
=- \int_{0}^{t}d\tau [\overline{\widetilde{\mathcal{H}}_{\text{fluc}}(0), [\widetilde{\mathcal{H}}_{\text{fluc}}(\tau)},\widetilde{\rho}(t)]] \label{eq:stationary}
\end{gather}
We also assume the existence of a bath correlation time $\tau_c$, such that the ensemble average becomes negligible for $\tau \gg \tau_c$. This allows us the extend the integration limit from $\int_{0}^{t} d\tau$ to $\int_{0}^{\infty} d\tau$ :
\begin{gather}
- \int_{0}^{\infty}d\tau [\overline{\widetilde{\mathcal{H}}_{\text{fluc}}(t), [\widetilde{\mathcal{H}}_{\text{fluc}}(t-\tau)},\widetilde{\rho}(t)]] \nonumber \\
= -\frac{1}{2} \sum_{m=-2}^{2} J_{\text{DD}}(m\omega_0) [T_{2m}(\vec{I_1},\vec{I_2}),[T_{2m}^{\dagger}(\vec{I_1},\vec{I_2}),\widetilde{\rho}(t)]] \nonumber \\
-\frac{1}{2} \sum_{j,k=1}^{2}\sum_{m=-1}^{1} J_{\text{CSA}}(m\omega_0) [T_{2m}(\vec{B},\vec{I_j}),[T_{2m}^{\dagger}(\vec{B},\vec{I_k}),\widetilde{\rho}(t)]] \nonumber \\
- \frac{1}{2}\sum_{j=1}^{2}\sum_{m=-1}^{1} J_{\text{CC}}(m\omega_0) ([T_{2m}(\vec{I_1},\vec{I_2}),[T_{2m}^{\dagger}(\vec{B},\vec{I_j}),\widetilde{\rho}(t)]] \nonumber \\
+  ([T_{2m}(\vec{B},\vec{I_j}),[T_{2m}^{\dagger}(\vec{I_1},\vec{I_2}),\widetilde{\rho}(t)]], \label{eq:redfield}
\end{gather}
with, assuming an exponentially decaying correlation function \cite{cavanagh1996protein,GOLDMAN2001160}, the spectral density functions read
\begin{gather}
J_{\text{DD}}(m\omega_0) = b^2\int_{-\infty}^{\infty} \expo{-imt} \,\expo{-\tau/\tau_c} d\tau = \frac{6 b^2}{5} \frac{2\tau_c}{1+(m\omega_0 \tau_c)^2}, \nonumber \\
J_{\text{CSA}}(m\omega_0) =\frac{6 d^2}{5} \frac{2\tau_c}{1+(m\omega_0 \tau_c)^2},~\text{and} \nonumber \\
J_{\text{CC}}(m\omega_0) =-\frac{6 bd}{5} \frac{2\tau_c}{1+(m\omega_0 \tau_c)^2}. \label{eq:cl_spec}
\end{gather}
So far we have not taken into account the temperature of the lattice with which the system spin is interacting so far, as a result the equation in its current form does not predict the correct thermal steady state. We follow the process described in \cite{BENGS2020106645} and 
\begin{itemize}
\item add correction \cite{EGOROV1998469,PhysRevLett.4.239} to the spectral densities so that they obey the detailed balance \cite{PhysRevLett.120.180502,PhysRevE.98.052104} : $J(\omega)\rightarrow \mathcal{J}(\omega) = \exp(-\beta \omega/2)J(\omega)$, where $\beta=1/K_B\,T$, and
\item replace all the double commutators of Eq.~\ref{eq:redfield} with respective Lindblad dissipator \cite{Annabestani2017,Karabanov18072014} as 
\begin{gather}
-\frac{1}{2}[A,[B,(\cdot)]] \rightarrow \mathcal{D}[A,B](\cdot) \nonumber \\ = A(\cdot)B - \frac{1}{2}(BA(\cdot) - (\cdot)BA). \nonumber
\end{gather}
\end{itemize}
After all these, the equation of motion for the evolution of the density matrix describing the full ensemble reads as Eq.~\Eqref{eq:GSKL} of the main text with
\begin{gather}
\Gamma_{\text{DD}}(\cdot) = \sum_{m=-2}^{2} \mathcal{J}_{\text{DD}}(m\omega_0)\, \mathcal{D}[T_{2m}(\vec{I_1},\vec{I_2}),T_{2m}^{\dagger}(\vec{I_1},\vec{I_2})] (\cdot) \nonumber \\
\Gamma_{\text{CSA}}(\cdot) = \sum_{j,k=1}^{2}\sum_{m=-1}^{1} \mathcal{J}_{\text{CSA}}(m\omega_0)\, \nonumber \\
 \mathcal{D}[T_{2m}(\vec{B},\vec{I_j}),T_{2m}^{\dagger}(\vec{B},\vec{I_k})] (\cdot), ~~\text{and} \nonumber \\
\Gamma_{\text{CC}}(\cdot) = \sum_{j=1}^2\sum_{m=-1}^{1} \mathcal{J}_{\text{CC}}(m\omega_0)\, (\mathcal{D}[T_{2m}(\vec{I_1},\vec{I_2}),T_{2m}^{\dagger}(\vec{B},\vec{I_j})] \nonumber \\
\mathcal{D}[T_{2m}(\vec{B},\vec{I_j}),T_{2m}^{\dagger}(\vec{I_1},\vec{I_2})])(\cdot) \label{eq:LB}
\end{gather}
Note that the equation is valid only for time $t$ satisfying Eq.~\ref{eq:COND} and the formal solution reads $\widetilde{\rho}(t) = \exp(\mathcal{L}t)\rho(0)$. As the generator $\mathcal{L}$, called the Liouvillian, is time independent, for a larger time the solution would read $\widetilde{\rho}(t) = \prod_{i=1}^{b}\exp(\mathcal{L}\Delta t)\rho(0)$, where $t=n\Delta t $ with each $\Delta t$ is such that Eq.~\ref{eq:COND} is satisfied. Since $\prod_{i=1}^{n}\exp(\mathcal{L}\Delta t) = \exp(\mathcal{L}\sum_{i=1}^{n}\Delta t) = \exp(\mathcal{L}t)$, the validity of Eq.~\Eqref{eq:GSKL} extends well beyond condition \Eqref{eq:COND}, thanks to the assumptions that lead to the time independence of $\mathcal{L}$. 

\subsection{C : Experimental Preparation of Initial States}

We start with the initial deviation density matrix $\rho_{\Delta}^{th} = 2(\rho^{th} - \mathbbm{1}/4)/\epsilon = I_{1z} + I_{2z}$. With the applications of the following pulses and a dephasing PFG in the end (as shown in Fig.\ref{fig:thseq}~(c)), we get 
\begin{gather*}
I_{1z} + I_{2z} \nonumber \\
\downarrow (\theta)_y \nonumber \\
\cos \theta\,(I_{1z} + I_{2z}) + \sin \theta (I_{1x} + I_{2x}) \nonumber \\
\downarrow 1/(2\Delta)~\mbox{with}~\Delta \gg J \nonumber \\
\cos \theta \, (I_{1z} + I_{2z}) + \sin \theta \, (-I_{1y} + I_{2y}) \nonumber \\
\downarrow (\theta)_{-x} - G_z \nonumber \\
I_{1z} + \cos (2\theta) I_{2z}.
\end{gather*}

\end{document}